\documentclass[aps,nofootinbib,amsmath,amsfonts,amssymb]{revtex4}
\usepackage{bm}

\def\bold#1{\bm{#1}}
\def\la{\langle}
\def\ra{\rangle}
\def\infint{\int_{-\infty}^\infty}
\def\piint{\int_{-\pi}^\pi}
\def\infsum#1{\sum_{{#1}=-\infty}^\infty}
\def\halfsum#1{\sum_{{#1}=0}^\infty}
\def\kappagroup{{\bold\kappa}}
\def\thetagroup{{\bold\theta}}
\def\Jkappa{J^{(\kappagroup)}}
\def\Ikappa{I^{(\kappagroup)}}

\def\Jtheta{J^{(\thetagroup)}}
\def\var{\text{var}}
\def\Pk#1{P({#1}|\kappagroup)}
\def\pk#1{p({#1}|\kappagroup)}

\begin{document}
\title{Fisher Information With Respect to Cumulants}

\author{S. Prasad}
\author{N. C. Menicucci}
\affiliation{Center for Advanced Studies and Department of Physics
and Astronomy\\ University of New Mexico\\ Albuquerque, New Mexico
87131}

\date{\today}

\begin{abstract}
Fisher information is a measure of the best precision with which a
parameter can be estimated from statistical data. It can also be
defined for a continuous random variable without reference to any
parameters, in which case it has a physically compelling
interpretation of representing the highest precision with which
the first cumulant of the random variable, i.e.,\ its mean, can be
estimated from its statistical realizations. We construct a
complete hierarchy of information measures that determine the best
precision with which all of the cumulants of a random variable --
and thus its complete probability distribution -- can be estimated
from its statistical realizations. Several properties of these
information measures and their generating functions are discussed.
\end{abstract}

\maketitle

\section{Introduction}
\label{sec:introduction}

Fisher information \cite{stats,fisherinfo} constitutes a central concept in statistical
estimation theory which furnishes a variety of useful estimates
of deterministic parameters from statistical observations typical of a
physical experiment. Its inverse yields a
lower bound, called the Cramer-Rao lower bound (CRLB), on the variance
of any unbiased estimator of a continuous parameter and thus limits
the best precision with which the parameter can be extracted from
statistical measurements \cite{fisherinfo,stats}.

It is useful to regard a probability density function (pdf) of a
random variable as being implicitly parameterized in terms of a
translational location parameter, e.g., its mean, median, or mode.
The Fisher information relative to such a purely translation
parameter is easily seen to be independent of that parameter, and
may be defined as the {\it Fisher information of the random
variable\/} \cite{fisherinfo} itself. The notion of Fisher
information of a random variable has been applied to the case in
which the random variable is the sample based mean. In the limit
of large sample size, asymptotic estimates \cite{fishersamplemean}
have been obtained in terms of the cumulants of the underlying pdf
and their derivatives.

In this correspondence, we present a further generalization of the Fisher
information of a continuous random variable.  Instead of using only
implicit parameterizations, we explicitly parameterize all
smooth, well-behaved pdf's in terms of their cumulants
\cite{mathworldcumulants,Kenney1951cumulants}.
Since the set of all cumulants of a pdf uniquely and
completely specifies the pdf, the Fisher information matrix relative
to all the cumulants should represent, in effect, the fidelity of
estimation of the full pdf from data.  The choice of cumulants to
parameterize a pdf is a particularly convenient one since, as we shall
see, it leads to a simple analytical form for the Fisher information
matrix. These concepts can be generalized still further, as we shall argue,
to the case of a discrete, integer-valued random variable.
We shall present some useful properties of these information measures,
discuss their generating functions, and illustrate our considerations
with simple examples.

\section{Fisher Information of a Continuous Random Variable}
\label{sec:fisherinfo}

Given a continuous random variable $X$ distributed according to the
pdf $p(x) > 0$ for all real $x$ (where $x$ is a statistical
realization of $X$), we may define a parameterized version of this
pdf, $p(x|\theta) = p(x - \theta)$.  Note that $p(x|\theta) = p(x)$ if
$\theta = 0$.  The Fisher information of $p(x|\theta)$ with respect to
$\theta$ is
\begin{eqnarray}
\label{eq:pdflocparam}
    J_\theta &=& \left\langle \left({\partial\ln p(x|\theta)\over \partial \theta}
    \right)^2 \right\rangle \nonumber \\
    &=& \infint dx\, p(x|\theta)\left({\partial\ln p(x|\theta)\over \partial \theta}
    \right)^2 \nonumber \\
    &=& \infint dx\, p(x-\theta)\left(-{\partial\ln p(x-\theta)\over \partial
    x}\right)^2 \nonumber \\
    &=& \infint {dx\over p(x)} \left({dp \over dx}\right)^2 \equiv J(X)\;,
\end{eqnarray}
where the angled brackets in the first line indicate expectation value
with respect to $p(x|\theta)$, and $J(X)$ is the Fisher information of
the random variable~$X$~\cite{fisherinfo}.  Note that because the pdf
is positive for all real values of $x$, we can integrate over the
infinite interval, and $J_\theta$ is manifestly independent of
$\theta$. Thus, any location parameter may be used for $\theta$, and
$J(X)$ is therefore a functional only of the shape of the
distribution, independent of its absolute location on any axis.

The Fisher information of a random variable $J(X)$ has two well-known
interpretations~\cite{stats,fisherinfo}. First, $J(X)$ quantifies the
statistical precision with which the location parameter $\theta$ of a
pdf on which the pdf depends translationally can be estimated from
data drawn according to the pdf $p(x-\theta)$.  On the other hand,
because $J(X)$ measures the mean squared slope of the log-likelihood
function $\ln p(x)$, it typically correlates with the narrowness of
the pdf or, equivalently, with the degree of statistical
reproducibility of the values assumed by the variable $X$. In the
Bayesian context, this narrowness is related to the extent of prior
knowledge about $X$. These two interpretations are related in that a
narrower pdf will provide higher ``resolution'' when used as a
measurement tool for determining the location parameter $\theta$.

A question of central interest in this correspondence is the following:
What information measures characterize the fidelity of a
statistical determination of the full pdf, not just its location
parameter? We shall see presently that a particularly simple
answer to this question can be obtained in terms of the Fisher
information matrix elements relative to the cumulants of the pdf
of the random variable.

\section{Cumulants of a PDF and Associated Fisher Information}
\label{sec:cumulants}

Every pdf $p(x)$ has a unique characteristic function associated
with it, given by its Fourier transform,
\begin{eqnarray}
\label{eq:charfxn}
    M(\nu) \equiv \la e^{i \nu x} \ra = \infint dx\, p(x)
    e^{i \nu x}\;,
\end{eqnarray}
which, in most cases, may be expressed in a power series in $\nu$
in terms of the moments of the pdf:
\begin{eqnarray}
\label{eq:charfxnmom}
    M(\nu) = \sum_{n=0}^\infty {(i\nu)^n\over n!}\mu_n'\;,
\end{eqnarray}
where $\mu_n' \equiv \la x^n \ra$ is the $n$th moment of the pdf
(about 0). Writing the logarithm of the characteristic function in
a similar series form defines the {\it cumulants} $\kappa_n$ of
the pdf as coefficients in the series expansion
\cite{mathworldcumulants,Kenney1951cumulants}
\begin{eqnarray}
\label{eq:lncharfxn}
    L(\nu) \equiv \ln M(\nu) \equiv \sum_{n=1}^\infty {(i\nu)^n\over
    n!}\kappa_n\;.
\end{eqnarray}
Note that the function $L(\nu)$ may be regarded as the generating
function for the cumulants, since
\begin{eqnarray}
\label{eq:cumgenfxn}
    \kappa_n=\left. {1\over i^n}{d^n L\over d\nu^n}\right|_{\nu=0}\;.
\end{eqnarray}
The cumulants $\kappa_n$ are related to the mean and central moments
of the pdf, the first few of them taking the form
$\kappa_1=\mu_1',\; \kappa_2=\mu_2,\; \kappa_3=\mu_3,\;
\kappa_4=\mu_4-2\mu_2^2,\; \ldots \;$, \
where $\mu_1'$ and $\mu_2$ are the mean and variance of the pdf, while
its third and fourth central moments, $\mu_3,\;\mu_4$, are related to
its skewness and kurtosis.  Since a pdf is related to its
characteristic function by a Fourier transform, we may thus
parameterize $p(x)$ in terms of the entire collection of its cumulants
(indicated by $\kappagroup$) as the following inverse Fourier transform:
\begin{eqnarray}
\label{eq:pdfparameterized}
    p(x) \rightarrow \pk{x} \equiv {1\over 2\pi}\infint
    d\nu\, e^{-i\nu x} \exp \left\{ \sum_{n=1}^\infty
    {(i\nu)^n\over n!}\kappa_n \right\} \;.
\end{eqnarray}

The Fisher information with respect to a set of real estimation
parameters $\thetagroup \equiv \{\theta_1, \theta_2, \ldots\}$ is
defined as a positive semi-definite matrix with elements
\cite{stats,fisherinfo}
\begin{eqnarray}
\label{eq:Jmatrixdef}
    \Jtheta_{mn} &\equiv& \left\langle
    {\partial \ln p(x|\thetagroup)\over \partial \theta_m}
    {\partial \ln p(x|\thetagroup)\over \partial \theta_n}
    \right\rangle\;,
\end{eqnarray}
where the expectation value is taken over the pdf, $p(x|\thetagroup)$,
of the statistical data from which the parameters are estimated. The
superscript $(\thetagroup)$ indicates the parameter set being used,
and both $n$ and $m$ range over the indices of the parameters in
$\thetagroup$. For a continuous pdf $p(x)$ that does not vanish at any finite value of $x$,
this becomes
\begin{eqnarray}
\label{eq:Jmatrixpdf}
    \Jtheta_{mn} = \infint {dx\over p(x|\thetagroup)}
    {\partial p(x|\thetagroup)\over \partial \theta_m}
    {\partial p(x|\thetagroup)\over \partial \theta_n}\;.
\end{eqnarray}
Noting from Eq.~(\ref{eq:pdfparameterized}) that
\begin{eqnarray}
\label{eq:dpdkappaeval}
    {\partial \pk{x}\over \partial \kappa_n} = {(-1)^n\over
    n!} {\partial^n \pk{x} \over \partial x^n}\;
\end{eqnarray}
and using Eq.~(\ref{eq:Jmatrixpdf}) with $\thetagroup$ chosen as the cumulant vector
$\kappagroup$, we obtain the Fisher information matrix relative
to the cumulants,
\begin{eqnarray}
\label{eq:CIMdef}
    \Jkappa_{mn}={(-1)^{m+n}\over m!n!}\infint {dx\over
    \pk{x}} {\partial^m \pk{x} \over \partial x^m}
    {\partial^n \pk{x} \over \partial x^n}\;,\quad\quad
    m,n=0,1,2,\ldots\quad.
\end{eqnarray}
While $\kappa_0$ is not a true cumulant,
including the possibility that $n$ and/or $m=0$ in Eq.~(\ref{eq:CIMdef})
leads formally to a particularly convenient generating function for these
elements, as we shall see in Sec.~\ref{sec:genfxn}.

Equation~(\ref{eq:CIMdef}) is the most important result of this
correspondence. It defines a complete hierarchy matrix of information measures,
which we may call the {\it cumulant information matrix} (CIM). The
diagonal elements of the inverse of the CIM yield the full hierarchy
of CRLB's on how precisely the various cumulants of $\pk{x}$ may be
estimated from the statistical realizations of $X$ \cite{stats} and
thus, in a sense, how well the entirety of the pdf may be estimated.

For $m=n=1$, the CIM element $\Jkappa_{11}$ reduces to $J(X)$ defined
in Eq.~(\ref{eq:pdflocparam}) as the Fisher information of the random
variable. This result confirms our earlier interpretation of $J(X)$ as
the precision with which a fiducial location parameter of the pdf may
be estimated.  Without knowledge of any higher order cumulants, it is
the first cumulant (the mean) that furnishes the most useful location
parameter of a pdf.

\section{A Generating Function for the Cumulant Information Matrix}
\label{sec:genfxn}

A useful technique for evaluating the CIM elements is the method of
generating functions. Upon multiplying both sides of
Eq.~(\ref{eq:CIMdef}) by $\lambda^m\mu^n$, then summing over all
non-negative integral values of $m$ and $n$, and finally interchanging
the order of integration and summations, a procedure justified by
the uniform convergence of the involved Taylor expansions,
we obtain the following result:
\begin{eqnarray}
\label{eq:genfxndef}
    \Jkappa(\lambda,\mu)&\equiv&\sum_{m,n=0}^\infty \Jkappa_{mn}
    \lambda^m\mu^n \\
\label{eq:genfxntaylor}
    &=&\infint {dx\over \pk{x}}
    \left[\sum_{m=0}^\infty{(-\lambda)^m\over m!}
    {\partial^m \pk{x} \over \partial x^m}\right]
    \left[\sum_{n=0}^\infty{(-\mu)^n\over n!} {\partial^n
    \pk{x} \over \partial x^n}\right]\;.
\end{eqnarray}
The two pairs of square brackets in Eq.~(\ref{eq:genfxntaylor})
enclose the Taylor expansions of $\pk{x-\lambda}$ and
$\pk{x-\mu}$, respectively.
We thus arrive at a rather simple form of the function $\Jkappa(\lambda,\mu)$:
\begin{eqnarray}
\label{eq:genfxn}
    \Jkappa(\lambda,\mu) = \infint dx\,{\pk{x-\lambda}
    \pk{x-\mu}\over \pk{x}}\;.
\end{eqnarray}
The function $\Jkappa(\lambda,\mu)$ given by Eq.~(\ref{eq:genfxn})
is a generating function for the CIM elements, since from
Eq.~(\ref{eq:genfxndef}) an arbitrary element $\Jkappa_{mn}$ may
be expressed as its partial derivative
\begin{eqnarray}
\label{eq:CIMfromgenfxn}
    \Jkappa_{mn} = {1\over m!n!}\left.{\partial^{m+n}
    \over \partial \lambda^m
    \partial \mu^n}\Jkappa(\lambda,\mu)\right|_{\lambda=\mu=0}\;.
\end{eqnarray}
The elements $\Jkappa_{0n} =
\Jkappa_{n0}$, as mentioned earlier, do not correspond to any information relative to any
cumulants, and can, in fact, be easily shown to vanish for $n \geq 1$,
while $\Jkappa_{00}=1$, the pdf normalization.

\section{Example: Gaussian pdf}
\label{sec:gaussian}

The Gaussian pdf provides an analytically tractable illustration of the
results of this correspondence.
For the Gaussian pdf,
\begin{eqnarray}
\label{eq:gaussian}
    p(x)={1\over \sqrt{2\pi\sigma^2}}e^{-(x-x_0)^2/2\sigma^2}\;,
\end{eqnarray}
its characteristic function is also Gaussian, and its cumulant generating
function is thus quadratic,
\begin{eqnarray}
\label{eq:charfxngauss}
    L(\nu)=\ln M(\nu) = i\nu x_0-\nu^2 \sigma^2/2.
\end{eqnarray}
The first two cumulants are thus
its mean $x_0$ and variance $\sigma^2$, while
all higher order cumulants vanish identically.  Parameterizing the
Gaussian in terms of its cumulants $\kappagroup$, we obtain, very
simply,
\begin{eqnarray}
\label{eq:paramgauss}
    \pk{x} = {1\over \sqrt{2\pi\kappa_2}}e^{-(x-\kappa_1)^2
    /2\kappa_2}\;.
\end{eqnarray}

The CIM generating function can be easily evaluated for the
Gaussian pdf (\ref{eq:paramgauss}), for which the integrand in
Eq.~(\ref{eq:genfxn}) is also Gaussian and easily integrated, with
the result
\begin{eqnarray}
\label{eq:genfxngauss}
    \Jkappa(\lambda,\mu) = e^{\lambda\mu / \kappa_2}
    = e^{\lambda\mu / \sigma^2}\;.
\end{eqnarray}
The individual CIM elements then follow from a use of
Eq.~(\ref{eq:CIMfromgenfxn}):
\begin{eqnarray}
\label{eq:CIMgauss}
    \Jkappa_{mn} = {\delta_{mn}\over n!\sigma^{2n}}\;.
\end{eqnarray}
The diagonal nature of the CIM
implies simple CRLB's on the variance of any set of unbiased estimators,
$\{\hat \kappa_n\}$, of the cumulants of the pdf:
\begin{eqnarray}
\label{eq:CRLBgauss}
    \var (\hat\kappa_n) \equiv \bigl\la (\hat \kappa_n-\kappa_n)^2
    \bigr\ra \geq n!\sigma^{2n}\;.
\end{eqnarray}
For general, biased estimators, the right-hand side of
Eq.~(\ref{eq:CRLBgauss}) must be replaced by the $nn-$diagonal
element of the matrix $B^T I B$, where the matrix $I$ is the
inverse of the CIM, which for the Gaussian case is diagonal with
the $nn-$element equal to $n!\sigma^{2n}$, $B$ is the bias matrix,
with elements $B_{mn}=\partial \la \hat \kappa_m\ra / \partial \kappa_n$,
and $B^T$ is its transpose \cite{stats}.

The $k$-statistics furnish useful unbiased estimators of the
cumulants of a distribution based on a finite sample drawn from that
distribution~\cite{mathworldk-stats,Kenney1962,Halmos1946}.  It is
well known that no cumulant estimator exists with a smaller variance than
that of the corresponding $k$-statistic.  
What our results show is that the $k$-statistics for a
Gaussian pdf are also {\it asymptotically efficient\/} estimators in
the sense that the CRLB~(\ref{eq:CRLBgauss}) is achieved in the limit
that the sample size $N \rightarrow \infty$. 
This asymptotic efficiency has not been previously derived for any of 
the $k$-statistics of order higher than 2. The variances of the
first few $k$-statistics for the Gaussian pdf are given below, along
with the associated CRLB's from the CIM analysis, the latter given as
the rightmost terms in the following expressions:
\begin{subequations}
\label{eq:comparegaussvark-stats}
\begin{eqnarray}
    \var(k_1) &=& {\kappa_2 \over N} \;=\; {\sigma^2 \over N}\\
    \var(k_2) &=& {2\kappa_2^2 \over N-1} \;\agt\; {2\sigma^4 \over
    N} \\
    \var(k_3) &=& {6N\kappa_2^3 \over (N-1)(N-2)} \;\agt\;
    {6\sigma^6 \over N} \\
    \var(k_4) &=& {24(N+1)N\kappa_2^4 \over (N-1)(N-2)(N-3)}
    \;\agt\; {24\sigma^8 \over N}\\
     &\vdots& \nonumber
\end{eqnarray}
\end{subequations}
(The CRLB for $N$ trials is $1/N$ times the CRLB for one trial.)
Notice that for the Gaussian, $k_1$ is an efficient estimator for any $N$,
and the others are asymptotically
efficient (indicated by $\agt$) in the limit $N \rightarrow \infty$.
This is true of all higher-order $k$-statistics for the Gaussian, as
well.

Two remarks are in order here. First, although any unbiased
estimator of a third or higher order cumulant is, on average, zero
for the Gaussian pdf, there is a finite statistical scatter in the
data from which the cumulants, regardless of their order, are
estimated. Second, the sharp increase of the CRLB's
(\ref{eq:CRLBgauss}) with increasing $n$ is a reflection of the
sharply decreased probability of occurrence of values of a
Gaussian variate in the wings of the pdf to which the higher order
cumulants are increasingly sensitive as a function of their order.
This is in fact a general result for any localized distribution;
the CIM elements $\rightarrow 0$ as $n,m \rightarrow \infty$,
because of the factor of $n!m!$ in the denominator of
Eq.~(\ref{eq:CIMdef}), resulting in CRLB's that increase without
bound as the order of the cumulant being estimated increases.

\section{An Inverse CIM Generating Function}
\label{sec:invgenfxn2}

The CIM elements are of value only insofar as they give a general
sense of how much information the data contain about the cumulants of
the pdf from which the data are drawn. It is the inverse of the CIM
that is needed to establish the CRLB, and such an inverse is often
difficult to calculate.  If a less stringent bound is acceptable, one
can use the reciprocal of the $nn$-diagonal element of the CIM
\cite{diagonalbound} to bound the minimum variance from below. Such a
bound, though easy to write down, is not optimal, however, since the
CRLB represents in general a greater lower bound.  The desire for a
simple way to calculate the inverse CIM motivates the following
attempt to define an appropriate generating function for the {\it
inverse\/} CIM and relate it to the CIM generating function defined in
Eq.~(\ref{eq:genfxn}).

Let us define the generating function $\Ikappa(\nu,\eta)$ for the
matrix elements $\Ikappa_{mn}$ of the inverse of the CIM as
\begin{eqnarray}
\label{eq:invgenfxn2}
    \Ikappa(\nu,\eta)\equiv\sum_{m,n=0}^\infty \Ikappa_{mn}{\nu^m
    \eta^n \over m!n!}\;,
\end{eqnarray}
from which the inverse CIM elements may be obtained as follows:
\begin{eqnarray}
\label{eq:invCIMfrominvgenfxn2}
    \Ikappa_{mn} = \left.{\partial^{m+n}
    \over \partial \nu^m
    \partial \eta^n}\Ikappa(\nu,\eta)\right|_{\nu=\eta=0}\;.
\end{eqnarray}
The factorials, $m!$ and $n!$, are included in the
definition~(\ref{eq:invgenfxn2}) to allow appropriate convergence
of the generating function (Compare Eq.~(\ref{eq:genfxndef})). We
may also define two ``marginal'' generating functions,
\begin{eqnarray}
\label{eq:marginals}
    \Jkappa_n(\lambda) \equiv \halfsum m \Jkappa_{mn} \lambda^m
    \qquad\text{and}\qquad \Ikappa_n(\nu) \equiv \halfsum \ell
    \Ikappa_{\ell n} {\nu^\ell \over \ell!}\;,
\end{eqnarray}
that are related to the full generating functions~(\ref{eq:genfxn})
and~(\ref{eq:invgenfxn2}) by
\begin{eqnarray}
\label{eq:marginaltofullgenfxn}
    \Jkappa_n(\lambda) = {1\over n!} \left.{\partial^n \over \partial
    \mu^n} \Jkappa(\lambda,\mu)\right|_{\mu=0} \qquad\text{and}\qquad
    \Ikappa_n(\nu) = \left.{\partial^n \over \partial
    \eta^n} \Ikappa(\nu,\eta)\right|_{\eta=0} \;.
\end{eqnarray}
By multiplying the two marginals (\ref{eq:marginals}), summing over the index $n$ from
0 to $\infty$, and noting that the symmetric matrices $\Ikappa_{mn}$ and $\Jkappa_{mn}$
are mutual inverses, we obtain the following relation:
\begin{eqnarray}
\label{eq:marginalgenfxnrelation2}
    \halfsum n \Jkappa_n(\lambda) \Ikappa_n(\nu)
    &=& \halfsum {k,m} \lambda^k {\nu^m \over m!} \halfsum n
    \Jkappa_{kn}\Ikappa_{mn} \nonumber \\
    &=& \halfsum {k,m} \lambda^k {\nu^m \over m!} \delta_{km}
    \nonumber \\
    &=& \vphantom{\halfsum n} e^{\lambda\nu}\;.
\end{eqnarray}
A Fourier analysis gives a second, integral relation,
\begin{eqnarray}
\label{eq:genfxnrelation2}
    {1\over 2\pi}\infint d\mu \infint d\eta\, e^{-i\mu\eta}
    \Jkappa(\lambda,i\mu) \Ikappa(\nu,\eta) =
    e^{\lambda\nu}\;,
\end{eqnarray}
as shown in the Appendix.  These relations are valid for
all values of $\lambda$ and $\nu$ in the complex plane.  In general,
relations~(\ref{eq:marginalgenfxnrelation2})
and~(\ref{eq:genfxnrelation2}) cannot be solved analytically. The case
of the Gaussian pdf is an exception,
for which the preceding relations can be solved and the generating function
$\Ikappa(\nu,\eta)$ derived exactly,
\begin{eqnarray}
\label{eq:gaussinvgenfxn2}
    \Ikappa(\nu,\eta)=e^{\nu\eta\,\sigma^2}\;.
\end{eqnarray}
Applying Eq.~(\ref{eq:invCIMfrominvgenfxn2}) to this result yields the
same CRLB's~(\ref{eq:CRLBgauss}) as obtained in
Sec.~\ref{sec:gaussian}.  While the CIM for the Gaussian is admittedly
trivial to invert, and thus it is not necessary to use the generating
function method in this case, we anticipate that in some cases it will
prove easier to approximate the solution of either
relation~(\ref{eq:marginalgenfxnrelation2})
or~(\ref{eq:genfxnrelation2}) than it would be to invert the full CIM.

\section{Generalization to a Discrete Random Variable}
\label{sec:discrete}

While the Fisher information of a continuously random variable
over the infinite interval may be defined through the use of an
arbitrary location parameterization, discrete distributions do not
readily permit an analogous definition. Although one cannot treat
a discrete random variable as a continuously-valued parameter, the
class of discrete distributions defined positive over the entirety
of the integers can always be parameterized in terms of their
cumulants and a corresponding CIM matrix defined.

For a discrete random variable $X$ distributed according to $P(x)$
with $x$ ranging over all integers, we may parameterize this
distribution as
\begin{eqnarray}
\label{eq:discreteparameterized}
    P(x) \rightarrow \Pk{x} \equiv {1\over 2\pi}\piint
    d\nu\, e^{-i\nu x} \exp \left\{ \sum_{n=1}^\infty
    {(i\nu)^n\over n!}\kappa_n \right\} \;,
\end{eqnarray}
where
\begin{eqnarray}
\label{eq:discretecumgenfxn}
    \kappa_n \equiv \left. {1\over i^n}{d^n \ln \la e^{i\nu x}\ra
    \over d\nu^n}\right|_{\nu=0}\;
\end{eqnarray}
are the cumulants in exact analogy to the continuous case. The only formal difference
is in terms of the limits of
integration in~(\ref{eq:discreteparameterized}), reflecting
the fact that $\la e^{i\nu x}\ra$ is now a Fourier {\it series}.\footnote{
For a random variable
that has a one-sided discrete realization over the set of all {\it non-negative}
integers only, the characteristic function $\langle \exp(i \nu x)\rangle$
is the $z$-transform \cite{Bracewell99} of the distribution $P(x)$, rather
than its Fourier series,
with $z=\exp(-i\nu)$. Notwithstanding this difference, the parameterization of
the distribution in terms of its cumulants is formally identical to that given in
Eq.~(\ref{eq:discreteparameterized}), and the considerations of this section apply
essentially unchanged.}

The parameterized distribution $\Pk{x}$ coincides with $P(x)$ for all
integral values of $x$, but it is also $C^\infty$ and
therefore defined for intermediate values of $x$ as well. These
intermediate values are not probabilities of anything, and they can
range outside the interval $[0,1]$. Still, because of its coincidence
with $P(x)$ at integral $x$, $\Pk{x}$ may be substituted for $P(x)$
when calculating an expectation value of any function of $X$. Most
importantly, though, we may now also take {\it derivatives\/} of this
parameterized distribution with respect to $x$, which are related to
partial derivatives with respect to the distribution's cumulants
exactly as in the continuous case, i.e.
\begin{eqnarray}
\label{eq:discretederiv}
    {\partial \Pk{x} \over \partial \kappa_n} = {(-1)^n \over n!}
    {\partial^n \Pk{x} \over \partial x^n}\;.
\end{eqnarray}
This allows us to define the discrete CIM elements exactly as in
the continuous case, with the infinite integral replaced by an
infinite sum:
\begin{eqnarray}
\label{eq:discreteCIMdef}
    \Jkappa_{mn} = {(-1)^{m+n}\over m!n!} \infsum x {1\over
    \Pk{x}} {\partial^m \Pk{x} \over \partial x^m}
    {\partial^n \Pk{x} \over \partial x^n}\;,\quad\quad
    m,n=0,1,2,\ldots.
\end{eqnarray}
The possibility that $n$ or $m = 0$ is also included in this
definition for assistance in defining a generating function for these
matrix elements, namely
\begin{eqnarray}
\label{eq:discretegenfxn}
    \Jkappa(\lambda,\mu) = \infsum x\,{\Pk{x-\lambda}
    \Pk{x-\mu}\over \Pk{x}}\;,
\end{eqnarray}
in exact analogy to the continuous case.  Without modification,
Eq.~(\ref{eq:CIMfromgenfxn}) may be used to generate the individual
CIM elements, and the generating function relations given in
Eqs.~(\ref{eq:marginalgenfxnrelation2}) and~(\ref{eq:genfxnrelation2})
also hold.

This is a powerful result.  While it is not possible to directly
define the Fisher information of a discrete random variable as in
the continuous case, we can define an analogous quantity: the
Fisher information with respect to the mean of the
cumulant-parameterized distribution. This is $\Jkappa_{11}$ of the
discrete CIM matrix. Since in the continuous case
$\Jkappa_{11}=J(X)$, we can {\it define\/} $J(X) \equiv
\Jkappa_{11}$ for a discrete distribution. The CIM matrix provides
a complete hierarchy of information measures with respect to the
cumulants in the discrete case, as well, and the inverse of the
CIM again gives the CRLB on any unbiased estimators of the
cumulants.

\section{Concluding Remarks}
\label{sec:conclusion}

The Fisher information of a continuous random variable can be
interpreted as the fidelity with which a fiducial location parameter
of a pdf (such as the mean) may be estimated from statistical data
drawn according to that distribution. In this correspondence, we have
introduced a more robust information measure---the cumulant
information matrix (CIM)---whose inverse bounds the variance of any
estimates of the cumulants of a pdf and, consequently, the fidelity
with which the entire pdf may be estimated. The Fisher information of
the random variable is included in this measure. We have also extended
the CIM concept to discrete random variables defined over integers,
for which the notion of Fisher information of the random variable is
ill defined.  We have also derived a generating function for the CIM
and given two relations between this generating function and a
generating function for the inverse CIM, which we hope will prove
useful in calculating CRLB's.  Further generalizations of this work
could include defining a cumulant information matrix or an analogous
quantity for multivariate distributions.  (See
McCullagh~\cite{McCullagh} for information on multivariate cumulants.)

We noted in Sections~\ref{sec:cumulants} and \ref{sec:discrete} that
the probability distribution to be parameterized must be strictly
positive. This was done to avoid singularities in the Fisher
information calculations.  As a general rule, the CIM for an arbitrary
random variable must be defined by restricting the interval of
integration or summation to the actual sample space of that
variable. This is particularly important when {\it a priori}
constraints like finiteness of support place restrictions on the set
of possible values of the allowed cumulant vectors.  One constraint
that is implicitly active in all of our calculations is that the
probability distribution, whether continuous or discrete, is
restricted to nonnegative values.  While this is obvious from the standpoint
of probability theory, it must be explicitly imposed on the space of all {\it estimable}
probability distributions. One only need consider the parameterization 
(\ref{eq:pdfparameterized}) to see why the cumulants $\kappagroup$ cannot take arbitrary
values if the left hand side of that equation is to remain nonnegative. 
CRLB's calculated in the present paper apply only to estimators that do not explore the
boundaries of the parameter space beyond which negative distributions are encountered. 
Incorporating edge effects will, in general,
reduce the minimum variance of a more general estimator, but such
bounds are not calculable from the Fisher information matrix since
inequality constraints do not affect its form.
However, certain other constraints such as support constraints can
alter the Fisher information matrix and change the method that one
must use to calculate CRLB's. The effects of constraints on CRLB's are
explored in detail by Gorman and Hero~\cite{Gorman1990}.

The considerations of this paper may be relevant to the general
area of inverse problems such as those concerning image
restoration from noisy image data. A particularly useful viewpoint
to adopt in discussions of image processing is to treat the
spatial distribution of intensity in an image, when properly
normalized, as representing the probability distribution of the
emission or detection of a photon over the image. From this
perspective, image restoration is equivalent to the problem of
estimating a probability distribution, the very problem we have
discussed here. The presence of noise in the actual image data
greatly compounds this estimation problem, a subject that requires
further study. A noteworthy algorithm which makes essential use of
this statistical viewpoint is the maximum entropy method
\cite{Narayan86}. Blind deconvolution methods \cite{Schulz93}
provide another example. They rely on the existence of constraints
to recover the point-spread function as well as the source
intensity distribution, both of which may be regarded as
appropriate pdf's.

\section*{Acknowledgments}

The work reported here was supported in part by the US Air Force
Office of Scientific Research under grants F49620-00-1-0155 and
F49620-01-1-0321.  We are also pleased to receive from Dr.~Chris Lloyd
a reprint of his paper cited here.

\appendix

\section{Derivation of A Generating Function Relation}

To prove Eq.~(\ref{eq:genfxnrelation2}), we first
note that
\begin{eqnarray}
\label{eq:Fourierxy}
    \infint dx \infint dy\; e^{-ixy} x^n y^m &=& \infint dx\; x^n
    \infint dy\; y^m e^{-ixy} \nonumber \\
    &=& \infint dx\; x^n i^m {\partial^m \over \partial x^m} \infint
    dy\; e^{-ixy} \nonumber \\
    &=& 2\pi i^m \infint dx\; x^n \delta^{(m)}(x)\;,
\end{eqnarray}
where $\delta^{(m)}(x)$ is the $m$th derivative of the Dirac
$\delta$-function. Integrating by parts $m$ times in the right-hand side of
Eq.~(\ref{eq:Fourierxy}) reduces that integral to $(-1)^m$ times
the $m$th derivative of $x^n$ evaluated at $x=0$, namely to $(-1)^m m!\delta_{mn}$,
and the following identity results:
\begin{eqnarray}
\label{eq:Kroneckerexpand}
    {1\over 2\pi} \infint dx \infint dy\; e^{-ixy} {(ix)^n \over n!}
    y^m &=& \delta_{mn}\;.
\end{eqnarray}
Now, from the generating function definitions~(\ref{eq:genfxndef})
and~(\ref{eq:invgenfxn2}) and from the marginal
definitions~(\ref{eq:marginals}), it is clear that
\begin{eqnarray}
\label{eq:genfxntomarginals}
    \Jkappa(\lambda,\mu) = \halfsum n \Jkappa_n(\lambda) \mu^n
    \qquad\text{and}\qquad \Ikappa(\nu,\eta) = \halfsum n
    \Ikappa_n(\nu) {\eta^n\over n!}\;.
\end{eqnarray}
Plugging these results into the left-hand side (LHS) of
Eq.~(\ref{eq:genfxnrelation2}), we obtain
\begin{eqnarray}
\label{eq:genfxnrelationderiv1}
    \text{LHS} &=& {1\over 2\pi}\infint d\mu \infint d\eta\, e^{-i\mu\eta}
    \left[\halfsum m \Jkappa_m(\lambda) (i\mu)^m \right]
    \left[\halfsum n \Ikappa_n(\nu) {\eta^n\over n!} \right]
    \nonumber \\
    &=& \halfsum {m,n} \Jkappa_m(\lambda)
    \Ikappa_n(\nu) \left[{1\over 2\pi}\infint d\mu \infint
    d\eta\, e^{-i\mu\eta} (i\mu)^m {\eta^n \over n!}\right]
    \nonumber \\
    &\overset{(a)}{=}& \halfsum {m,n} \Jkappa_m(\lambda)
    \Ikappa_n(\nu)\; \delta_{mn}
    \nonumber \\
    &\overset{(b)}{=}& \vphantom{\halfsum n} e^{\lambda\nu}
\end{eqnarray}
where Eq.~(\ref{eq:Kroneckerexpand}) was used in step (a), and
Eq.~(\ref{eq:marginalgenfxnrelation2}) allowed step (b).  Thus,
relation~(\ref{eq:genfxnrelation2}) is established.


\begin{thebibliography}{10}

\bibitem{stats} H.~L. Van Trees, {\it Detection, Estimation, and
Modulation Theory\/}, Part I. New York: John Wiley, 1968.

\bibitem{fisherinfo}T.~M. Cover and J.~A. Thomas, {\it Elements of Information
Theory\/}. New York: John Wiley, 1991, Sec.~12.11.

\bibitem{fishersamplemean} C.~J. Lloyd, ``Asymptotic expansions of the
Fisher information in a sample mean,'' {\it Statistics \& Probability
Lett.\/}, vol.~11, pp.~133-137 (1991).

\bibitem{mathworldcumulants}
E.~W. Weisstein. (1999). Cumulant. [Online]. Available:
http://mathworld.wolfram.com/Cumulant.html

\bibitem{Kenney1951cumulants}J.~F. Kenney and E.~S. Keeping,
``Cumulants and the Cumulant-Generating Function,'' ``Additive
Property of Cumulants,'' and ``Sheppard's Correction,'' in {\it
Mathematics of Statistics\/}, 2nd ed., Pt. 2. Princeton, NJ: Van
Nostrand, 1951, \S4.10--4.12, pp.~77--82.

\bibitem{mathworldk-stats}
E.~W. Weisstein. (1999). $k$-statistic. [Online]. Available:
http://mathworld.wolfram.com/k-Statistic.html

\bibitem{Kenney1962}
J.~F. Kenney and E.~S. Keeping, ``The $k$-Statistics,'' in
{\it Mathematics of Statistics}, 3rd ed., Pt. 1. Princeton, NJ:
Van Nostrand, 1962, \S7.9, pp. 99--100.

\bibitem{Halmos1946} P.~R. Halmos, ``The theory of unbiased
estimation,'' {\it Ann.\ Math.\ Stat.\/}, vol. 17, pp.~34--43 (1946).

\bibitem{diagonalbound} H.~H. Barrett, J.~L. Denny, R.~F. Wagner, and
K.~J. Myers, ``Objective assessment of image quality. II. Fisher
information, Fourier crosstalk, and figures of merit for task
performance,'' {\it J.~Opt.~Soc.~Am. A\/}, vol. 12, pp.~834-852 (1995).

\bibitem{Bracewell99}
R. Bracewell, {\it The Fourier Transforms and Its Applications\/}, 3rd ed., New York, NY:
McGraw Hill, 1999, pp.~257-262.

\bibitem{Gorman1990} J.~D. Gorman and A.~O. Hero, ``Lower bounds for
parameer estimation with constraints,'' {\it IEEE\ Trans.\ Inf.\
Theory\/}, vol. 26, pp.~1285-1301 (1990).

\bibitem{Narayan86} R. Narayan and R. Nityananda, ``Maximum Entropy Restoration
in Astronomy,'' {\it Ann.~Rev.~Astron.~Astrophys.\/}, vol. 24, pp.~127-170 (1986).

\bibitem{Schulz93} T.~J. Schulz, ``Multi-frame Blind Deconvolution of Astronomical
Images,'' {\it J.~Opt.~Soc.~Am. A\/}, vol. 10, pp.~1064-1073 (1993).

\bibitem{McCullagh} P.~McCullagh, {\it Tensor Methods in
Statistics\/}.  New York: Chapman and Hall, 1987.

\end{thebibliography}
\end{document}